# Correlations ia a polymeric structure immersed in a magnetic solution

**B. V. Costa**

Laboratório de Simulação, Departamento de Física, ICEx
Universidade Federal de Minas Gerais, 31720-901 Belo Horizonte, Minas Gerais, Brazil

E-mail: bvc@fisica.ufmg.br

**Abstract.**
Polymers are among the most important materials in the modern society being found almost in every activity of our daily life. Understanding their chemical and physical properties lead to improvements of their usage. The correlation functions are one of most important quantities to understand a physical system. The characteristic way it behaves describe how the system fluctuates, and much of the progress achieved to understand complex systems has been due to their study. Of particular interest in polymer science are the space correlations which describe its mechanical behavior. In this work I study the stiffness of a polymer immersed in a magnetic medium and trapped in an optical tweezers. Using Monte Carlo simulations the correlation function along the chain and the force in the tweezers are obtained as a function of temperature and density of magnetic particles. The results show that the correlation decay has two regimes: an initial very fast decay of order the monomer-monomer spacing and a power law in the long distance regime. The power law exponent has a minimum at a temperature $T_{min}$ for any non zero density of magnetic particles indicating that the system is more correlated in this region. Using a formula for the persistence length derived from the WLC theory one observed that it has a maximum at the same temperature. These results suggest that the correlations in the system may be a combination of exponential and power law.

## 1. Introduction

Polymers are a class of complex molecules exhibiting interesting and important properties that can respond to external stimuli as, for instance, temperature, solvent composition and external magnetic field [1, 2]. The astonishing achievements along the last decades with a large variety of applications elects the polymer research as one of the most important fields for technological applications as well as presenting several fundamental theoretical challenges. The development of magnetic materials based on polymers with magnetic inclusions are important in a wide range of applications covering a broad area with a tremendous potential from drug targeting in medicine, braille reading, devices to heat generators and even to remove unwanted flavour compounds from wine [3–5]. In the presence of a solution rich in magnetic particles the polymer changes its mechanical properties which in turn can be tuned through a magnetic field [6, 7].

The Worm Like Chain (WLC) [8], sometimes referred to as the Kratky-Porod model, is a general model used to describe the behavior of semi-flexible polymers. In the WLC model the polymer is understood as successions of $N$ rigid segments of length $b$ oriented in the direction $\vec{u}$. The energy of a given configuration is the sum of bending energies corresponding to neighbour segments. For a chain with $N$ monomers its energy is written as

$$E_{WLC} = -\frac{J}{b}\sum_{i=1}^{N} \vec{u}_i \cdot \vec{u}_{i+1} = -\frac{J}{b}\sum \cos\theta_i \quad . \tag{1}$$

The WLC, described by Eq. 1 corresponds to the one dimensional Heisenberg model [9] in magnetism which is known to have no phase transition at any finite temperature, $T$. In the thermodynamic limit the correlation function decays exponentially as $C(s = |i-j|) = \langle \vec{u}_i \cdot \vec{u}_j \rangle \sim exp(-bs/\xi)$, where $\xi = J/k_BT$ is the correlation length. In the polymer language it is known as the persistence length, measuring the distance over which the bonds are correlated along the polymer chain. In other words, it characterizes the system's stiffness. Another important quantity is the gyration radius, $R_g$, defined in terms of the chain end-to-end distance as

$$R_g^2 = (b\sum_{i=1}^{N} \vec{u}_i)^2 \approx 2Nb\xi = 2L_0\xi \quad , \quad (2)$$

where $L_0 = Nb$ is the chain length. Measuring $R_g$ gives an estimate of the persistence length. A more precise way to estimate $\xi$ is provided by stretching the polymer and measuring the force acting in the chain. Let $f\hat{y}$ be an external force acting on the extremities of the system in the $\hat{y}$ direction. The energy due to the force adds a new term in equation Eq. 1 as

$$E^*_{WLC} = -\frac{J}{b}\sum_{i=1}^{N} \vec{u}_i \cdot \vec{u}_{i+1} = -\frac{J}{b}\sum \cos\theta_i - Fb\sum_{i=1}^{N} \cos\phi_i \quad , \quad (3)$$

where $\varphi_i$ is the angle between the bond $\vec{u}_i$ and $\hat{y}$. This energy corresponds to the Heisenberg model in an external magnetic field that, unfortunately, has no exact analytical solution. An approximate interpolation formula for the persistence length was obtained by Marko and Siggia and improved by Bustamante [10, 11]

$$f = \frac{k_BT}{\xi}\{x - \frac{1}{4} + \frac{1}{4(1-x)^2} + \sum_{i=2}^{7} a_i x^i\} \quad (4)$$

with $a_2 = -0.5164228, a_3 = -2.737418, a_4 = 16.07497, a_5 = -38.87607, a_6 = -39.49944$ and $a_7 = -14.17718$. This equation is asymptotically exact in the large and small-force limits. Atomistic simulation is an invaluable toll in many branches of science. With the advances in computing power driven by the development of new processors and algorithms a computational approach enable predictions and provide explanations of experimentally observed phenomena and opens up possibilities for the development of new analytical models [12]. The aim in this work concerns the use of computer simulations to study the mechanical properties of polymers when diluted in a solvent rich in magnetic particles. Even if the polymer does not contains any magnetic monomer, when in the presence of a magnetic solution some magnetic particles can stick in the polymeric chain changing its mechanical properties. The physical properties of the polymer can be studied by measuring the force exerted by the polymer in an optical trap [11, 13, 14].

I will consider the polymer stretched by optical tweezers in a solution of magnetic particles. The polymer is simulated by employing the generic model of a flexible elastic homopolymer where non-bonded interactions are represented by Lennard-Jones (LJ) potential and adjacent monomers in the linear chain interact via the standard anharmonic FENE (finitely extensible nonlinear elastic) potential [15–17]. Solvent-monomer interact through LJ forces. The solventsolvent particles interact via dipole and LJ potentials. Due to the simplicity of the WLC model it is not expected to describe in full the present simulation [18–21]. The results presented in this work show two distinct regimes of the persistence length. Initially, a fast exponential decay is observed for short distances and a power law decay in the long distance limit. Although the correlation function does not have a pure exponential behavior, it is common to define an effective persistence length by fitting the experimental data to Eq. 4. Surprisingly, the exponential decay and the power law describe qualitatively the same mechanical behavior of the system.

This work is organized in the following way. In section 2 I define the model, describe the setup used and give some details about the simulation method. In section 3 the simulation results are presented and in section 4 I present my conclusions and some final remarks.

## 2. The Model and Simulation Details

To model the polymer I choose an atomistic point of view. As usual the monomers are considered as spherical beads interacting with each other through Lennard-Jones (6-12) potential and the anharmonic FENE (finitely extensible nonlinear elastic) potential [12, 15, 22, 23]. The LennardJones pair potential is given by

$$V_{LJ} = -4\epsilon \left[ \left(\frac{\sigma}{r_{ij}}\right)^6 - \left(\frac{\sigma}{r_{ij}}\right)^{12} \right], \quad (5)$$

where $r_{ij}$ is the distance between particles labeled $i$ and $j$. The potential has a minimum at $r_{min} = 2^{1/6}$ with $V_{LJ}(r_{min}) = V_{min} = -\epsilon$. The FENE interaction is defined by the following formula

$$V_{FENE} = -\frac{1}{2} K R^2 \ln\left[1 - \left(\frac{r_{min} - r_{ij}}{R}\right)\right]. \quad (6)$$

The maximum bond extension is limited by the FENE potential, which diverges at $r_{ij} \rightarrow r_{min} \pm R$. The polymer is immersed in a fluid containing a number of magnetic dipolar particles that interact through Lennard-Jones and magnetic dipole ($V_{Dipole}$) potentials [24].

$$V_{Dipole} = -D \left\{ \frac{\left(\vec{S}_i \cdot \vec{r}_{i,j}\right)\left(\vec{S}_j \cdot \vec{r}_{i,j}\right)}{|\vec{r}_{ij}|^5} - \frac{\vec{S}_i \cdot \vec{S}_j}{|\vec{r}_{ij}|^3} \right\}, \quad (7)$$

where $\vec{S}_j$ is the magnetic moment of particle $j$. The total energy of the system is given by

$$E^*_{Total} = V^{m-m}_{LJ} + V^{m-m}_{FENE} + V^{d-m}_{LJ} + V^{d-d}_{Dipole}, \quad (8)$$

where $m-m$, $m-d$ and $d-d$ represent the interactions monomer-monomer, monomer-magnetic particles and between magnetic particles respectively. The polymer is attached at one end to a rigid substrate and the other is trapped on optical tweezers [13, 14] as shown schematically in Fig. 1. Because the presence of the tweezers an additional term accounting for this must be included in the total energy Eq. 8. In general the optical tweezers force can be described by a harmonic oscillator potential, $V_{Tweezer} = 1/2 k \delta r^2$, where the constant $k$ is the trap stiffness and $\delta r$ is the deviation from the equilibrium position. The final expression for the total energy becomes $E = E_{Total^*} + V_{Tweezer}$. In figure 1 it is shown the main elements of the simulation setup. The system is contained in a cubic box of volume $2L_x \times L_y \times 2L_z$. Rigid walls boundary conditions were used in all directions. Along the simulation distance will be measured in units of $r_{min}$ the energy in units of and temperature in units of $\epsilon/k_B$. To time evolute the system I used the Metropolis algorithm. When referred, the time interval is to be understood as a Monte Carlo step (MCS). A MCS consists in an attempt to move all particles and sweep all magnetic moments in the system using the Metropolis algorithm [25]. Special care was taken with pseudo-random number generator used to prevent spurious correlations [26, 27]. Here, it was used the *xoshift*1024** pseudo-random number generator implemented in *gfortran-2008*. This generator has a period of $2^{256}-1$. As initial conditions the dipole particles were distributed at random inside the simulation box. The first phase of the simulation is used to equilibrate the system. By monitoring the energy, the system was considered in equilibrium when the averaged energy does not change for more than 0.1% after a number $N_{Eql}$ of MCS. $10^5$ independent initial configurations are generated for a particular set of parameters, then, for each such configurations $10^4$ measurements are made. It is a common strategy to use a cutoff in the Lennard-Jones and Dipole potentials to speedup the code execution. In the present case I decided not to use this strategy to keep the simulation as trustworthy as possible. The price paid was that I could not use very large systems sizes. Having this in mind a fixed size polymer with $N = 200$ was used. To the initial conditions the $N$ monomers were distributed along the $y$ direction with coordinates ($x = \delta x, y = nr_{min} + \delta y, z = \delta z$), with $n = 2,3...N$. Here, $\delta\alpha$ ($\alpha = x,y,z$) are small random fluctuation. In every case a simulation at low temperature ($T \leq 0.01$) is done to estimate the polymer size at the ground state $L$. For each configuration are calculated the force exerted by the tweezes in the last bead and the orientated correlation, C, of bond vectors along the chain. The appropriate correlation function for polymer is defined in the following way. Let $\sim u_i$ to be the vector pointing from monomer $i - 1$ to $i$. The correlation along the polymer chain is

$$C(s) = \langle \cos\theta(s) \rangle = \frac{\langle \vec{u}_i \cdot \vec{u}_{i+s} \rangle}{\langle \vec{u}_i \rangle} \qquad (9)$$

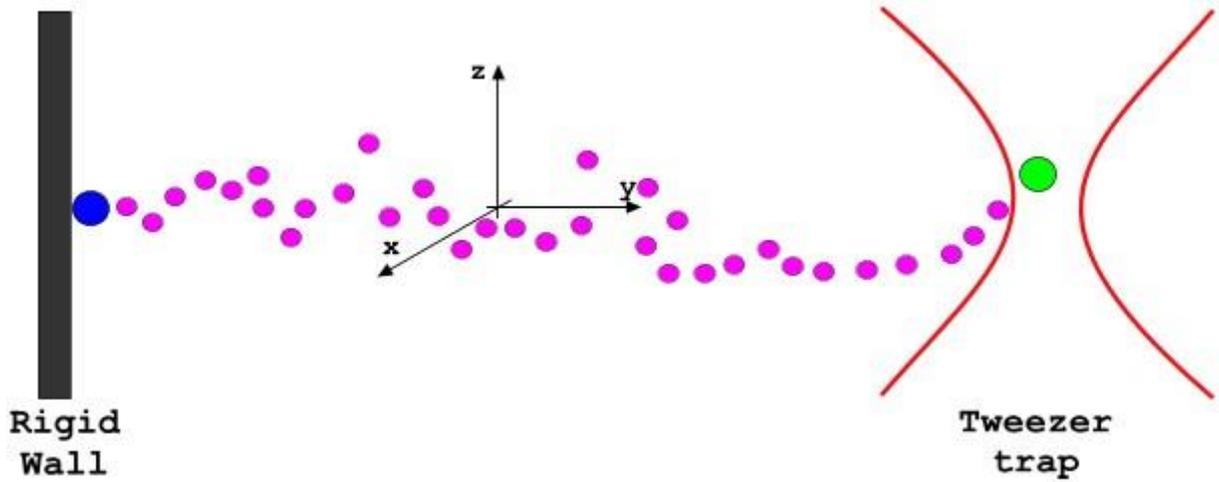

Figure 1: (Color online) Setup of the computational experiment. The polymer is represented as violet beads. One end of the polymer (Blue bead) is attached to a rigid wall, the other extreme is trapped in an optical tweezers (Green). For clarity, the solvent is not shown.

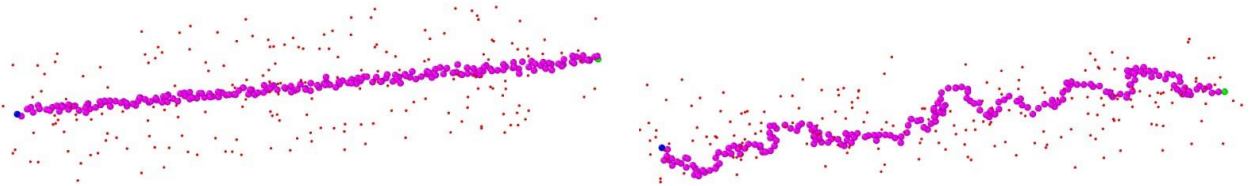

Figure 2: (Color online) Snapshots of the initial condition (left) and for the in equilibrium system after $10^5$ MCS. There are 200 monomers in the polymer (Violet) and 100 magnetic particles in the solution (Small red). The blue particle is fixed at $y = 0$ and the green particle is in the optical trap.

From the point of view of the Worm-Like Chain (WLC) model it is possible to associate C to the persistence length, $\xi$, as $C = \exp\left(\frac{-s\langle \vec{u}_i \rangle}{\xi}\right)$. However, it is not expected that this behavior can be extracted from the present system due to electrostatics and long-range interactions not included in the WLC model. HP Hsu et al. [18] using lattice models have pointed out that equation 9 "is not true for real polymers, irrespective of the considered conditions". Besides, the presence of the rigid wall in one end and the tweezers in the other break the symmetry introducing an additional source of discrepancy. Due to this lack of symmetry I will consider the correlation taking the monomer attached to the rigid wall ($y = 0$) as the reference monomer.

## 3. Results

In figure 3 it is shown the simulation results for a polymer with 200 monomers       For the simulations for all $T$ and $\rho$ values one observed the short range correlations decay rapidly. An attempt (Not shown here) to adjust an exponential decay to the simulated data gives a persistence length less than $\frac{r_{min}}{2}$. Out of this region the correlation decay is slower indicating a power law behavior. In figure 4 it is shown the exponent, $\eta$, obtained by adjusting a function $C(s) = A + B(1 - s/L)^\eta$ to the correlations $C(s)$ with a density $\rho = 0.5 \times 10^{-3}$ and $10^{-2}$ of magnetic particles. Here, density is define as the ratio of the number of magnetic particles to the box volume ($V = 2L_x \times L_y \times 2L_z$). One observe that for $\rho = 0$ (Red squares in Fig. 4) the exponent $\eta$ increases monotonically as a function of temperature while for $\rho > 0$ a minimum is observed at an intermediate temperature, $0.2 < T_{max} < 0.25$. The minima seems to occur at the same temperature for both concentrations. A possible interpretation for this behavior is as follows. At low temperature the magnetic particles attach to the polymer in an almost fixed position. The attraction between those particles induces a folding of the polymer chain (See Fig. 2) diminishing the correlations with a consequent increase of the exponent $\eta$. As temperature augments it is unlikely that the magnetic particles remain closely attached to the polymer approximately recovering the $\rho = 0$ behavior.

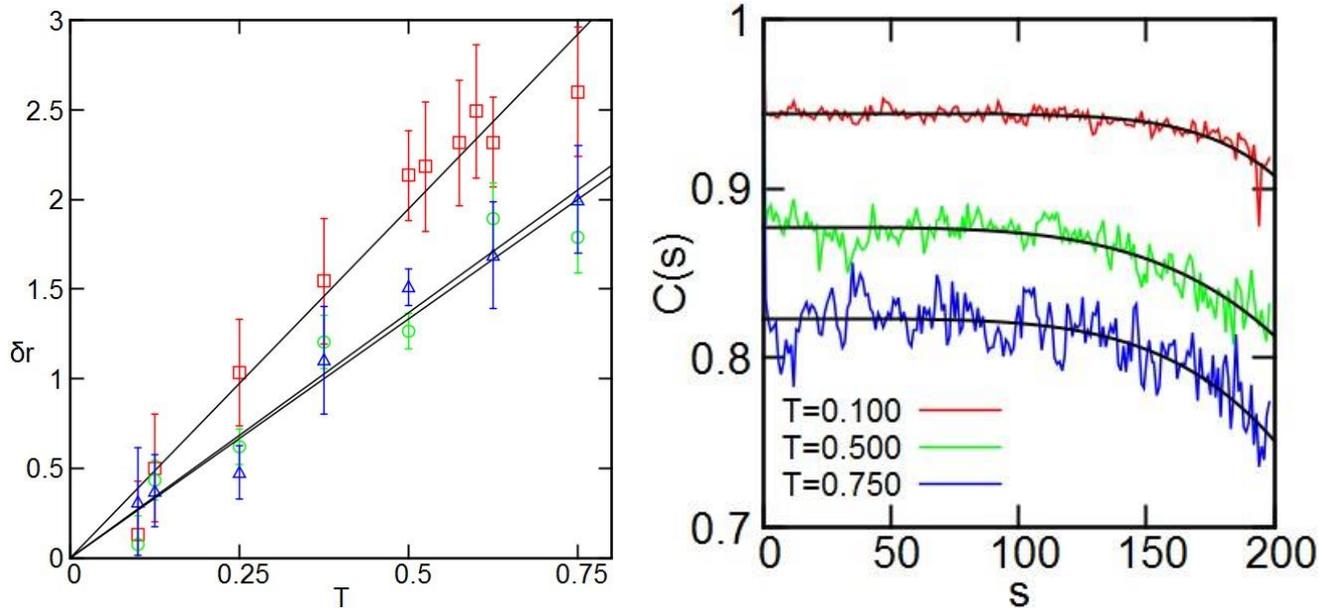

Figure 3: (Color online) The left hand sid plot shows the distance to the equilibrium point of the last bead inside the optical trap as a function of temperature for a polymer with 200 nonmagnetic monomers. The red, green and blue points are for 0,100 and 200 magnetic particles in the solvent, respectively. The straight line was obtained using the best fit of a straight line $\delta r = aT$ to the simulated points. The rhs is a plot for the correlation function as defined in equation 9. The red, green and blue curves are the simulated results. The black lines are adjusts using $C(s) = A + B(1 - s/L)^\eta$. Vertical lines indicate the region where $\chi$ has a minimum.

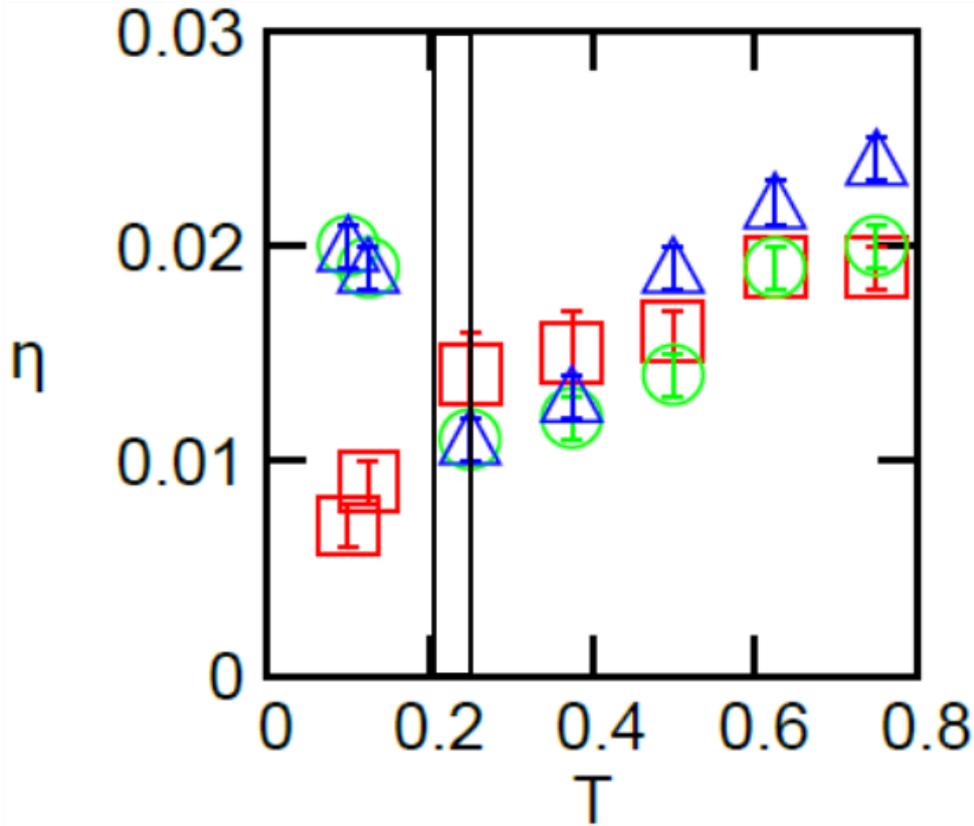

Figure 4: (Color online)

The force exerted by the tweezers in the polymer is easily calculated in the harmonic limit by using the displacement of the last bead from the equilibrium position, as shown in Fig. 3. Using the expression for the persistence length described by the WLC model (Equation 4) and the results for the force exerted by the tweezers ($\frac{1}{2}k(\delta r)^2$) an effective persistence length can be obtained as shown in figure 5. As should be expected the persistence length decays monotonically with temperature for $\rho = 0$. However, for $\rho > 0$ a maximum develops at a temperature that coincides with the region were the exponent $\eta$ has a minimum. Both approaches agree in describing qualitatively the correlations behavior. This odd result indicates that the models studied by HP Hsu et al. [18] are of restricted applicability when considering continuous models. Another possibility is that the correlations have a hybrid behavior between exponential and power law behavior.

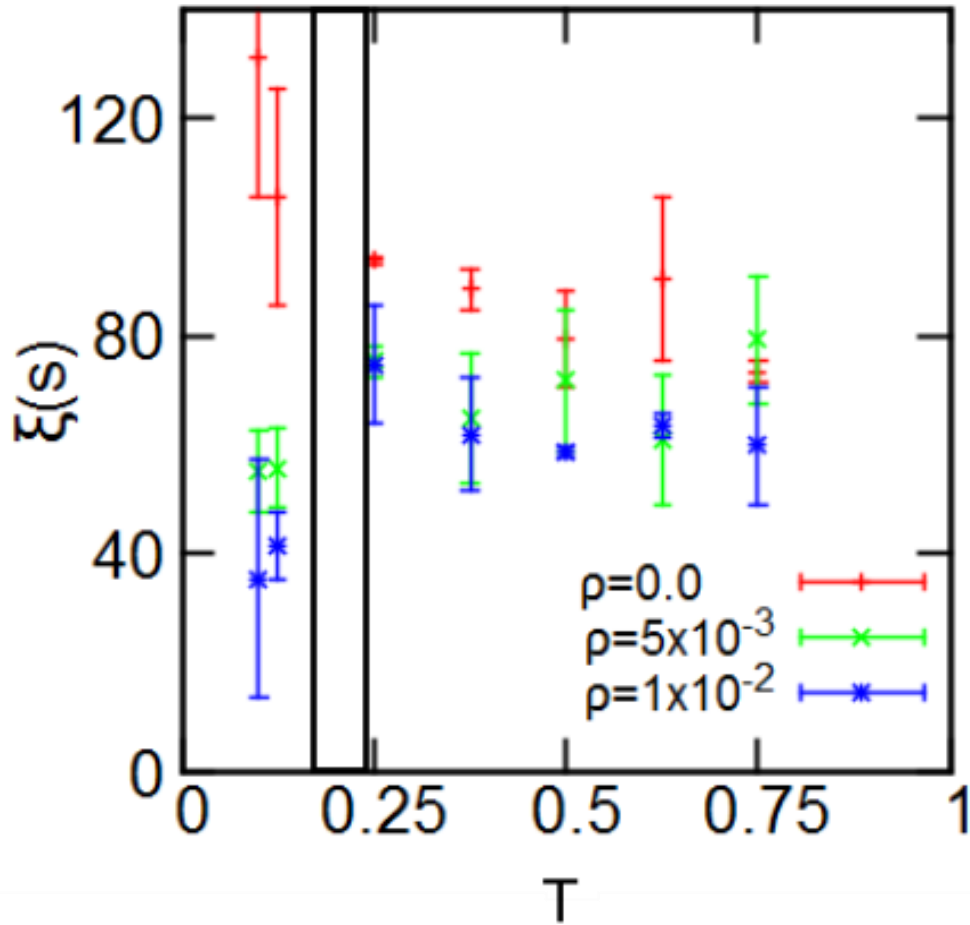

Figure 5: (Color online) Persistence length as a function of temperature calculated using the WLC model (4). The magnetic particle density is shown in the inset. Vertical lines indicate the region where $\chi$ has a maximum.

### 4. Final Remarks

The correlation functions are one of most important quantities to understand a physical system. The characteristic way it behaves describe how the system fluctuates, and much of the progress achieved to understand complex systems has been due to their study. Of particular interest in polymer science are the space correlations which describe its mechanical behavior. In the celebrated WLC model it is expected that the correlations have an exponential decay with a characteristic persistence length. However, due to the simplicity of the model it is not expected to describe more complex situations. In this work I used Monte Carlo simulations to study the mechanical behavior of a polymer stretched by an optical tweezers immersed in a fluid with magnetic particles. The results show that the correlation decay has two regimes: an initial with a very fast decay of order of the monomer spacing and a power law in the long distance regime. The power law exponent monotonically diminish with temperature in the absence of magnetic particles ($\rho = 0$), however it has a minimum at a temperature $T_{min}$ for ($\rho \neq 0$) indicating that the system is more correlated in this region. An effective persistence length can be obtained by using a formula derived from the WLC theory. For $\rho = 0$ it decreases monotonically, but showing a maximum for $\rho \neq 0$ at the same temperature $T_{min}$. Those results may indicate that the correlations in the system may be a combination between exponential and power law. To confirm this hypotheses a more systematic study is necessary, however those calculations are very time consuming being beyond our computational resources at the moment. An interesting subject would be to obtain the relationship between the persistence length and the density of magnetic particles in the medium.

**Acknowledgments**

This work was partially supported by CNPq and Fapemig, Brazilian Agencies. BVC thanks CNPq and FAPEMIG for the support under grants CNPq 402091/2012-4 and FAPEMIG RED-00458-16.

**References**

[1]  Bower D I 2003 An introduction to polymer physics

[2]  Philippova O 2000 *POLYMER SCIENCE SERIES CC/C OF VYSOKOMOLEKULIARNYE SOEDINENIIA* **42** 208–228

[3]  Philippova O, Barabanova A, Molchanov V and Khokhlov A 2011 *European polymer journal* **47** 542–559

[4]  Mosbach K and Schr¨oder U 1979 *FEBS letters* **102** 112–116

[5]  Liang C, Jeffery D W and Taylor D K 2018 *Molecules* **23** 1140

[6]  Kalia S, Kango S, Kumar A, Haldorai Y, Kumari B and Kumar R 2014 *Colloid and Polymer Science* **292** 2025–2052

[7]  Wilson J, Poddar P, Frey N, Srikanth H, Mohomed K, Harmon J, Kotha S and Wachsmuth J 2004 *Journal of Applied Physics* **95** 1439–1443

[8]  Theodorakopoulos N 2019 *Statistical Physics Of Dna: An Introduction To Melting, Unzipping And Flexibility Of The Double Helix* (World Scientific)

[9]  Heisenberg W 1985 Zur theorie des ferromagnetismus *Original Scientific Papers Wissenschaftliche Originalarbeiten* (Springer) pp 580–597

[10] Marko J F and Siggia E D 1995 *Macromolecules* **28** 8759–8770

[11] Bouchiat C, Wang M D, Allemand J F, Strick T, Block S and Croquette V 1999 *Biophysical journal* **76** 409–413

[12] Gartner III T E and Jayaraman A 2019 *Macromolecules* **52** 755–786

[13] Williams M C 2002 *Biophysics Textbook Online: http://www. biophysics. org/btol*

[14] Wang M D, Yin H, Landick R, Gelles J and Block S M 1997 *Biophysical journal* **72** 1335

[15] Bird R, Armstrong R and Hassager O 1987 *New York: Wiley* **2** 2–2

[16] Kremer K and Grest G S 1990 *The Journal of Chemical Physics* **92** 5057–5086

[17] Milchev A, Bhattacharya A and Binder K 2001 *Macromolecules* **34** 1881–1893

[18] Hsu H P, Paul W and Binder K 2010 *Macromolecules* **43** 3094–3102

[19] Wittmer J, Meyer H, Baschnagel J, Johner A, Obukhov S, Mattioni L, Mu¨ller M and Semenov A N 2004 *Physical review letters* **93** 147801

[20] Wittmer J, Beckrich P, Meyer H, Cavallo A, Johner A and Baschnagel J 2007 *Physical Review E* **76** 011803

[21] Shirvanyants D, Panyukov S, Liao Q and Rubinstein M 2008 *Macromolecules* **41** 1475–1485

[22] Oobatake M and Ooi T 1972 *Progress of theoretical physics* **48** 2132–2143

[23] Koci T, Qi K and Bachmann M 2016 The impact of bonded interactions on the groundstate geometries of a small flexible polymer *Journal of Physics: Conference Series* vol 759 p 012013

[24] Brydson J A 1999 *Plastics materials* (Elsevier)

[25] Landau D P and Binder K 2014 *A guide to Monte Carlo simulations in statistical physics* (Cambridge university press)

[26] Resende F J and Costa B V 1998 *Physical Review E* **58** 5183

[27] Ferrenberg A M, Landau D P and Wong Y J 1992 *Physical Review Letters* **69** 3382